\documentstyle[aps,prb,twocolumn,epsf,floats]{revtex}
\draft
\begin{document}
\wideabs{

%%%%%%%%%%%%%%%%%%%%%%%%%%%%%%%%%%%%%%%%%%%%%%%%%%%%%%%%%%%%%%%%%%%%%%
\title{Photoluminescence from fractional quantum Hall systems: \\
       Role of separation between electron and hole layers}
%%%%%%%%%%%%%%%%%%%%%%%%%%%%%%%%%%%%%%%%%%%%%%%%%%%%%%%%%%%%%%%%%%%%%%

\author{
   Arkadiusz W\'ojs}
\address{
   Department of Physics, 
   University of Tennessee, Knoxville, Tennessee 37996 \\
   Institute of Physics, 
   Wroclaw University of Technology, Wroclaw 50-370, Poland}

\author{
   John J. Quinn}
\address{
   Department of Physics, 
   University of Tennessee, Knoxville, Tennessee 37996}

\maketitle

\begin{abstract}
The photoluminescence (PL) spectrum of a two-dimensional electron gas 
(2DEG) in the fractional quantum Hall regime is studied as a function 
of the separation $d$ between the electron and valence hole layers.
The abrupt change in the response of the 2DEG to the optically injected 
hole at $d$ of the order of the magnetic length $\lambda$ results in
a complete reconstruction of the PL spectrum.
At $d<\lambda$, the hole binds one or two electrons to form neutral 
($X$) or charged ($X^-$) excitons, and the PL spectrum probes the 
lifetimes and binding energies of these states rather than the original 
correlations of the 2DEG.
At $d>2\lambda$, depending on the filling factor $\nu$, the hole 
either decouples from the 2DEG to form an ``uncorrelated'' state 
$h$ or binds one or two Laughlin quasielectrons (QE) to form 
fractionally charged excitons $h$QE or $h$QE$_2$.
The strict optical selection rules for bound states are formulated, 
and the only optically active ones turn out to be $h$, $h$QE* (an 
excited state of the dark $h$QE), and $h$QE$_2$.
The ``anyon exciton'' $h$QE$_3$ suggested in earlier studies is 
neither stable nor radiative at any value of $d$.
The critical dependence of the stability of different states on the 
presence of QE's in the 2DEG explains the observed anomalies in the 
PL spectrum at $\nu={1\over3}$ and ${2\over3}$.
\end{abstract}
\pacs{71.35.Ji, 71.35.Ee, 73.20.Dx}
}

\section{Introduction}
\label{secI}
%%%%%%%%%%%%%%%%%%%%%%%%%%%%%%%%%%%%%%%%%%%%%%%%%%%%%%%%%%%%%%%%%%%%%%
%%%%%%%%%%%%%%%%%%%%%%%%%%%%%%%%%%%%%%%%%%%%%%%%%%%%%%%%%%%%%%%%%%%%%%
The optical properties of quasi-two-dimensional (2D) electron systems 
in high magnetic fields have been extensively studied in the recent 
years both experimentally\cite{heiman,turberfield,goldberg,buhmann1,%
goldys,kukushkin,takeyama,gravier,pinczuk,kheng,buhmann2,shields1,%
finkelstein,hayne,nickel,tischler,wojtowicz,jiang,brown,kim} and 
theoretically.\cite{lerner,dzyubenko1,macdonald1,macdonald2,wang,%
apalkov,rashba,chen,bilayer-e,stebe,x-dot,palacios,x-fqhe,x-cf,x-pl,%
whittaker,pawel1,pawel2}
In symmetrically doped quantum wells (QW), where both conduction 
electrons and valence holes are confined in the same 2D layer, the 
photoluminescence (PL) spectrum of an electron gas (2DEG) probes the
binding energy and optical properties of neutral and charged excitons 
(bound states of one or two electrons and a hole, $X=e$--$h$ and 
$X^-=2e$--$h$), rather than the original correlations of the 2DEG 
itself.
The experiments\cite{kheng,buhmann2,shields1,finkelstein,hayne,nickel,%
tischler,wojtowicz,jiang,brown,kim} and theory\cite{stebe,x-dot,%
palacios,x-fqhe,x-cf,x-pl,whittaker} agree that the $X^-$ can exist in 
the form of a number of different bound states, whose binding energies 
depend strongly on the well width $w$ and composition, magnetic field 
$B$, etc., but (at least in dilute systems) much less on the electron 
filling factor $\nu$.
In particular, the only bound $X^-$ state that occurs at zero or low $B$ 
is the optically active singlet\cite{stebe,x-pl,whittaker} $X^-_s$, while 
more bound states form at higher $B$.
Of these states, one is observed in PL,\cite{shields1,finkelstein,hayne,%
kim} and it has only recently been identified\cite{x-pl} as an excited 
``bright'' triplet $X^-_{tb}$.
The lowest energy ``dark'' triplet $X^-_{td}$ has been predicted 
earlier,\cite{x-dot} but it is expected to have very long optical 
lifetime\cite{palacios} and its recombination has not yet been 
detected experimentally.\cite{hayne,x-pl,whittaker} 

The PL spectra containing more information about the original electron 
correlations of the 2DEG are obtained in asymmetrically doped wide QW's
or heterojunctions, where the spatial separation $d$ of electron and 
hole layers weakens the $e$--$h$ interaction.\cite{macdonald2}
Unless $d$ is smaller than the magnetic length $\lambda$, the PL spectra 
of such bi-layer systems show no recombination from $X^-$ states.
Instead, they show anomalies\cite{heiman,turberfield,goldberg,buhmann1,%
goldys,kukushkin,takeyama} at the filling factors $\nu={1\over3}$ and 
${2\over3}$ at which Laughlin incompressible fluid states\cite{laughlin} 
are formed in the 2DEG and the fractional quantum Hall (FQH) effect
\cite{tsui} is observed in transport experiments.

The present paper is a continuation of our earlier work\cite{bilayer-e}
where we studied the energy spectra of 2D fractional quantum Hall systems 
in the presence of an optically injected valence hole.
There, we have identified the possible bound states in which a valence 
hole can occur.
They included the ``uncorrelated'' state $h$ in which the free hole 
moves in the rigid electron Laughlin fluid at a local filling factor 
$\nu={1\over3}$, and the fractionally charged exciton (FCX) states $h$QE 
and $h$QE$_2$ in which the hole binds one or two Laughlin quasielectrons 
(QE).
The charge neutral ``anyon exciton'' state\cite{rashba} $h$QE$_3$ was 
found unstable at any value of $d$.
Here, we give a detailed anaysis of the optical properties of these 
states and explain the features observed in the PL spectra of bi-layer 
systems.
Based on the analysis of the involved dynamical symmetries\cite{avron,%
dzyubenko2} (those of charged particles moving in a translationally 
invariant space and in a perpendicular magnetic field) we formulate the 
optical selection rules for the FCX complexes.
These rules are verified in exact numerical calculations for finite 
systems in Haldane's spherical geometry\cite{haldane2,wu} (using 
Lanczos-based algorithms\cite{lanczos} we are able to calculate the 
exact spectra of up to nine electrons and a hole at $\nu\approx{1\over3}$).
It turns that the only radiative bound states involving the hole are 
$h$, $h$QE* (an excited state of $h$QE), and $h$QE$_2$, and that emission 
from both $h$QE and $h$QE$_3$ is forbidden.
The fact that the previously suggested\cite{macdonald2} recombination 
from a $h$--QE pair state can only occur through the excited state 
$h$QE* diminishes the importance of this process at low temperatures.
The result that the $h$QE$_3$ complex (or any of its excitations) is 
neither stable nor radiative questions applicability of the theory of 
``anyon excitons'' put forward by Rashba and Portnoi\cite{rashba} to 
account for the anomalies observed in the PL spectra at $\nu={1\over3}$ 
and ${2\over3}$.
Instead, these anomalies are explained in terms of emission from the
competing {\em radiative} bound states, $h$QE* and $h$QE$_2$, and 
from an uncorrelated hole state $h$.

\section{Model}
\label{secII}
%%%%%%%%%%%%%%%%%%%%%%%%%%%%%%%%%%%%%%%%%%%%%%%%%%%%%%%%%%%%%%%%%%%%%%
%%%%%%%%%%%%%%%%%%%%%%%%%%%%%%%%%%%%%%%%%%%%%%%%%%%%%%%%%%%%%%%%%%%%%%
The model considered here is identical to that of 
Ref.~\onlinecite{bilayer-e}.
A 2DEG in a strong magnetic field $B$ fills a fraction $\nu<1$ of 
the lowest LL of a narrow QW, whose width $w$ we set to zero.
A small number ($\nu_h\ll\nu$) of valence holes are optically injected 
into a parallel 2D layer of width $w_h=0$, separated from the electron 
layer by a distance $d$. 
The single-particle states $\left|m\right>$ in the lowest LL are the 
eigenstates of the orbital angular momentum, $m=0$, $-1$, $-2$, \dots\ 
for the electrons and $m_h=-m=0$, 1, 2, \dots\ for the holes.
Since $\nu_h\ll\nu$ and the strongly bound complexes at large $B$ 
involve only one hole, it is enough to study the many-electron--one-hole 
Hamiltonian which can be written as
\begin{equation}
\label{eq1}
  H = \sum_{ijkl} \left( c_i^\dagger c_j^\dagger c_k c_l V^{ee}_{ijkl}
                + c_i^\dagger h_j^\dagger h_k c_l V^{eh}_{ijkl} \right),
\end{equation}
where $c_m^\dagger$ ($h_m^\dagger$) and $c_m$ ($h_m$) create and 
annihilate an electron (hole) in state $\left|m\right>$.
The constant energy of the lowest LL is removed from $H$, which hence 
includes only the $e$--$e$ and $e$--$h$ interactions whose two-body 
matrix elements $V^{ee}$ and $V^{eh}$ are defined by the intra- and 
inter-layer Coulomb potentials, $V_{ee}(r)=e^2/r$ and $V_{eh}(r)=
-e^2/\sqrt{r^2+d^2}$.
At $d=0$, the $e$--$h$ matrix elements are equal to the $e$--$e$ 
exchange ones, $V^{eh}_{ijkl}=-V^{ee}_{ikjl}$, and at $d>0$ the 
$e$--$h$ attraction at short range is reduced.
The convenient units of length and energy are the magnetic length 
$\lambda$ and the energy $e^2/\lambda$.

The 2D translational invariance of $H$ results in the conservation of 
two orbital quantum numbers: the projection of total angular momentum 
${\cal M}=\sum_m(c_m^\dagger c_m-h_m^\dagger h_m)m$ and an additional 
angular momentum quantum number ${\cal K}$ associated with partial 
decoupling of the center-of-mass motion of an $e$--$h$ system in a 
homogeneous magnetic field.\cite{avron,dzyubenko2}
For a system with a finite total charge, ${\cal Q}=\sum_m(h_m^\dagger 
h_m-c_m^\dagger c_m)e\ne0$, the partial decoupling of the center-of-mass 
motion means that the energy spectrum consists of degenerate LL's.
\cite{avron} 
The states within each LL are labeled by ${\cal K}=0$, 1, 2, \dots\ 
and all have the same value of ${\cal L}={\cal M}+{\cal K}$.
Since both ${\cal M}$ and ${\cal K}$ (and hence also ${\cal L}$) commute 
with the PL operator ${\cal P}$, which annihilates an optically active 
(zero-momentum, $k=0$) $e$--$h$ pair (exciton), ${\cal M}$, ${\cal K}$, 
and ${\cal L}$ are all simultaneously conserved in the PL process.

The 2D symmetry of a planar system is preserved in the finite size,
$N$-electron--one-hole ($Ne$--$h$) calculation in Haldane's geometry,
\cite{haldane2} where all particles are confined to a spherical 
surface of radius $R$ and the radial magnetic field is produced by 
a Dirac monopole.
The conversion of the numerical results between the spherical and 
planar geometries follows from the exact mapping\cite{x-pl,geometry} 
between the planar quantum numbers ${\cal M}$ and ${\cal K}$ and 
the 2D algebra of the total angular momentum ${\bf L}$ on a sphere.
The detailed description of the Haldane sphere model can be found 
elsewhere\cite{haldane2,wu,fano} (see also Refs.~\onlinecite{x-fqhe,%
x-cf,x-pl} for the application to $e$--$h$ systems).
The strength $2S$ of the magnetic monopole is defined in the units 
of flux quantum $\phi_0=hc/e$, so that $4\pi R^2B=2S\phi_0$ and the 
magnetic length is $\lambda=R/\sqrt{S}$.
The single-particle states are the eigenstates of angular momentum 
$l\ge S$ and its projection $m$, and are called monopole harmonics.
The single-particle energies fall into degenerate angular momentum 
shells (LL's).
The lowest shell has $l=S$, and thus $2S$ is a measure of the system 
size through the LL degeneracy.
The charged many-electron--one-hole states form degenerate total 
angular momentum ($L$) multiplets (LL's) of their own.
The total angular momentum projection $L_z$ labels different states of 
the same multiplet just as ${\cal K}$ or ${\cal M}$ did for different 
states of the same LL on a plane.
Different multiplets are labeled by $L$ just as different LL's on 
a plane were labeled by ${\cal L}$.
The pair of optical selection rules on the sphere, $\Delta L_z=\Delta 
L=0$ (equivalent to $\Delta{\cal M}=\Delta{\cal K}=0$ on a plane) 
results from the fact that an optically active exciton has zero angular 
momentum.

\section{Bound States}
\label{secIII}
%%%%%%%%%%%%%%%%%%%%%%%%%%%%%%%%%%%%%%%%%%%%%%%%%%%%%%%%%%%%%%%%%%%%%%
%%%%%%%%%%%%%%%%%%%%%%%%%%%%%%%%%%%%%%%%%%%%%%%%%%%%%%%%%%%%%%%%%%%%%%
\subsection{Small Layer Separation}
\label{secIII_1}
%%%%%%%%%%%%%%%%%%%%%%%%%%%%%%%%%%%%%%%%%%%%%%%%%%%%%%%%%%%%%%%%%%%%%%
Depending on the separation $d$ between the electron and hole layers,
different bound states can occur in an $e$--$h$ system.
\cite{bilayer-e}
In the ``strong coupling'' regime, at $d$ less than about $1.5\lambda$,
the interaction between the hole and the electrons is stronger than 
the characteristic correlation energy of the 2DEG.
The response of the 2DEG to the optically injected valence hole occurs 
through spontaneus creation of charge excitations (QE--QH pairs) which 
completely screen its charge. 
As a result, the original $e$--$e$ correlations of the 2DEG are locally 
(in the vicinity of the hole) replaced by (stronger) $e$--$h$ correlations.
These new correlations are most conveniently described in terms of 
two types of new quasiparticles formed in the system, neutral ($X$) 
or charged ($X^-$) exciton states, in which the hole binds one or two 
electrons, respectively.

In an ideal system with no LL mixing and $w=w_h=d=0$, the ``dark'' 
(non-radiative; see Sec.~\ref{secIV}) triplet charged exciton 
$X^-_{td}$ is the only bound state (other than the neutral exciton 
$X$) that is stable in the presence of the surrounding 2DEG
\cite{x-dot,palacios,x-pl} (e.g., $X^-_2+e\rightarrow2X^-$ for the 
charged bi-exciton).
The $X^-_{td}$ unbinds\cite{bilayer-e,palacios} at $d\approx\lambda$, 
and a different $X^-$ state, a dark singlet $X^-_{sd}$, forms
\cite{bilayer-e,fox} at $0.4\lambda\le d\le1.5\lambda$.
In more realistic systems, when the effects due to LL mixing, finite 
widths of electron and hole layers, and their finite separation are 
taken into account, a few other bound $X^-$ states occur.\cite{x-pl} 
Most important of these states are the bright singlet $X^-_s$ and 
the bright triplet $X^-_{tb}$.
The four $X^-$ states are distinguished by the total electron spin 
$J$, and the total angular momentum $L$: the $X^-_s$, $X^-_{tb}$, 
$X^-_{td}$, and $X^-_{sd}$ states have $J=0$, 1, 1, 0, and $L=S$, 
$S$, $S-1$, $S-2$, respectively (on a plane, $L=S$, $S-1$, and $S-2$ 
correspond to ${\cal L}=0$, $-1$, and $-2$, respectively).
The binding energies of $X^-$ states depend strongly on $B$ and $d$.
In a narrow ($w\approx w_h\approx10$~nm) and symmetric ($d=0$) GaAs QW,
the $X^-_s$ is the most strongly bound $X^-$ state at $B$ smaller than 
about 30~T, and at larger $B$ the $X^-$ ground state changes to the 
$X^-_{td}$ (the $X^-_{tb}$ has always smaller binding energy than 
both $X^-_s$ and $X^-_{td}$).
At $d>0$, the binding energy of the $X^-_s$ is reduced more than 
that of the two triplets, and the critical value of $B$ at which 
the singlet--triplet crossing occurs is significantly decreased.
\cite{izabela}  

It has not been clearly spelled out until recently\cite{bilayer-e,review} 
that only the ``decoupled''\cite{lerner,dzyubenko1,macdonald1} $k=0$ 
state of the charge neutral $X$ exists in the 2DEG.
The $X$ at $k>0$ has a finite electric dipole moment (proportional
to $k$) whose strong interaction with the underlying 2DEG leads to 
the binding of the second electron and the formation of an $X^-$.
The numerical calculations show\cite{bilayer-e,review} that the low 
lying band of $e$--$h$ states at $L>0$, previously interpreted
\cite{macdonald1,wang,apalkov} as the dispersion of a charge neutral 
``dressed exciton'' (an $X$ with $k>0$ coupled to the QE--QH pair 
excitations of the 2DEG), in fact describe an $X^-$. 

\begin{figure}[t]
\epsfxsize=3.40in
\epsffile{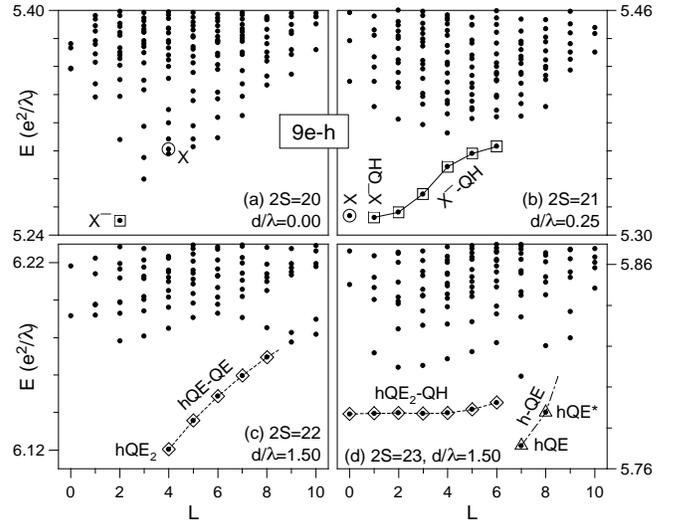}
\caption{
   The energy spectra (energy $E$ vs.\ angular momentum $L$) of 
   the $9e$--$h$ system calculated on a Haldane sphere with different 
   monopole strengths $2S$ and at for different layer separations $d$:
   (a) $2S=20$ and $d=0$, 
   (b) $2S=21$ and $d=0.25\lambda$, 
   (c) $2S=22$ and $d=1.5\lambda$, 
   (d) $2S=23$ and $d=1.5\lambda$.
   Different symbols and lines mark states and bands containing
   different quasiparticles:
   circles -- $X$, 
   squares -- $X^-$, 
   diamonds -- $h$QE$_2$, 
   triangles -- $h$QE and $h$QE*. 
   $\lambda$ is the magnetic length.}
\label{fig1}
\end{figure}

As an example, in Fig.~\ref{fig1}(ab) we present the $9e$--$h$ energy 
spectra at $2S=20$ and 21 and at small layer separation ($d=0$ and 
$0.25\lambda$, respectively), in which the lowest energy states have 
been identified as containing an $X^-$ state (this is the ``dark'' 
triplet state $X^-_{td}$ bound in the lowest LL) Laughlin correlated 
with the remaining seven electrons and an $X$ decoupled from the 
remaining eight electrons.
While a detailed discussion of these states has been given elsewhere,
\cite{bilayer-e} let us only note that the low energy band of states 
connected with a line in Fig.~\ref{fig1}(b) describes an $X^-_{td}$ 
interacting with a QH of the two-component $e$--$X^-_{td}$ incompressible 
fluid with Laughlin correlations\cite{laughlin,halperin} and not the
``dressed exciton'' dispersion (the angular momenta $L=1$, 2, \dots, 
6 of this band can be predicted from a generalized, two-component 
$e$--$X^-$ composite fermion picture\cite{x-cf}).

It is quite remarkable that the wavefunction of the 2DEG in the strong 
coupling regime can be well represented in terms of wavefunctions of 
competing bound $X$ and $X^-$ states, neglecting the distorsion of 
these states due to the coupling to the surrounding electrons.
For the $X^-$ states, this is a consequence of the short range
\cite{parentage} of the $e$--$X^-$ repulsion which results in Laughlin 
correlations and the effective isolation\cite{x-pl} of the $X^-$ states 
from the 2DEG.
For the $X$ state at $k=0$ (whose charge and electric dipole moment 
vanish), this is a result of weak (zero at $d=0$) coupling to the 2DEG.

\subsection{Large Layer Separation}
\label{secIII_2}
%%%%%%%%%%%%%%%%%%%%%%%%%%%%%%%%%%%%%%%%%%%%%%%%%%%%%%%%%%%%%%%%%%%%%%
At $d$ larger than about $1.5\lambda$, in the ``weak coupling'' regime, 
the $e$--$h$ attraction becomes too weak compared to the characteristic 
$e$--$e$ correlation energy, its range becomes too large compared to the 
characteristic $e$--$e$ separation, and the $X$ and $X^-$ states unbind.
In this regime, the perturbation associated with the potential of the 
optically injected hole does not cause the reconstruction of the $e$--$e$ 
(Laughlin) correlations of the 2DEG, whose response involves only the 
existing Laughlin QE's.
Since no additional QE--QH pairs are spontaneously created to screen
the hole charge, a discontinuity occurs at the Laughlin fillings such
as $\nu={1\over3}$.
At $\nu\le{1\over3}$, no QE's that could bind to the hole occur in the 
2DEG, the existing QH's are repelled from it, and the electrons in the 
vicinity of the hole form a Laughlin state with the (local) filling 
factor $\nu={1\over3}$.
In this ``uncorrelated'' state, the hole causes no (local) response 
of the 2DEG. 
At $\nu>{1\over3}$, the hole binds one or two QE's to form fractionally
charged excitonic states $h$QE or $h$QE$_2$ (it has been shown
\cite{bilayer-e} that the charge neutral ``anyonic excitons'' $h$QE$_3$ 
are unstable at any value of $d$).
Just as in the case of the $X$ or $X^-$ at small $d$, the $h$QE$_n$ 
states are well defined quasiparticles of the $e$--$h$ system at larger 
$d$, and they can be attributed such single particle properties as the 
binding energy $\Delta$, angular momentum $L$, PL energy $\omega$ and 
oscillator strength $\tau^{-1}$, etc.
Because of their low density, the $h$QE$_n$ quasiparticles can be to 
a good approximation regarded as non-interacting, free particles moving 
in a ``rigid'' Laughlin $\nu={1\over3}$ reference state.

While the binding of $h$QE$_n$ states as a function of $d$ and $\nu$ 
has been discussed in great detail in Ref.~\onlinecite{bilayer-e}, 
in Fig.~\ref{fig1}(cd) we present the $9e$--$h$ energy spectra at 
$2S=22$ and 23 and at $d=1.5\lambda$, in which the lowest energy 
states contain the $h$QE, $h$QE* (the first excited state of the 
$h$--QE pair), and $h$QE$_2$ complexes.
As for the $X$ and $X^-$ states in the strong coupling regime, it is
quite remarkable that the complicated correlations of a many body 
$e$--$h$ system at larger $d$ can be well represented in terms of 
rather simple and well defined free $h$QE$_n$ quasiparticles.

\section{Optical Selection Rules}
\label{secIV}
%%%%%%%%%%%%%%%%%%%%%%%%%%%%%%%%%%%%%%%%%%%%%%%%%%%%%%%%%%%%%%%%%%%%%%
%%%%%%%%%%%%%%%%%%%%%%%%%%%%%%%%%%%%%%%%%%%%%%%%%%%%%%%%%%%%%%%%%%%%%%
A number of different selection rules govern the optical recombination 
of bound $e$--$h$ complexes.
In general, any symmetry resulting in a conservation of a certain
quantum number ${\cal W}$ results in a strict selection rule 
\begin{equation}
   \Delta{\cal W}={\rm const}
\end{equation}
if the commutator between ${\cal W}$ and the PL operator 
\begin{equation}
   {\cal P}=\sum_m(-1)^m c_m h_m
\end{equation}
which annihilates an optical ($k=0$) exciton (on a Haldane sphere) 
is proportional to ${\cal P}$.

The so-called ``hidden symmetry''\cite{lerner,dzyubenko1,macdonald1} 
is the exact particle--hole symmetry in the lowest LL of narrow QW's 
in which electrons and valence holes are confined to the same layer 
(equal widths, $w=w_h$, and zero separation, $d=0$).
As a result, the optically active ($k=0$) excitons annihilated by 
${\cal P}$ decouple from the excess electrons.
The $k=0$ exciton is the only radiative bound state of an $e$--$h$ 
system, and the emission from the so-called ``multiplicative'' (MP) 
many-body states which contain a number ($N_X$) of $k=0$ excitons 
occurs at the bare exciton energy (independent of the electron density)
and follows the $\Delta N_X=-1$ selection rule.
Because of the exciton decoupling, all bound states other than the 
exciton (e.g., the triplet charged exciton $X^-_{td}$) have $N_X=0$ 
and cannot recombine.
The hidden symmetry holds only to some extent in realistic systems, 
where the asymmetry ($w\ne w_h$) and separation ($d>0$) of electron 
and hole layers, as well as the asymmetric LL mixing (due to different
electron and hole cyclotron energies) result, for example, in the 
binding of the radiative singlet ($X^-_s$) and triplet ($X^-_{tb}$) 
charged exciton states.

The 2D translational/rotational symmetry results in the conservation 
of ${\cal M}$ and ${\cal K}$ (or $L_z$ and $L$ on a sphere) in the 
emission process.\cite{x-fqhe,x-cf,x-pl,dzyubenko2}
The $\Delta{\cal M}=\Delta{\cal K}=0$ (or $\Delta L_z=\Delta L=0$) 
selection rules hold strictly when applied to the entire $e$--$h$ 
system or to an isolated bound state.
These rules (independently from the $\Delta N_X=-1$ rule) forbid 
emission from the $X^-_{td}$ state which has ${\cal L}={\cal M}+
{\cal K}=-1$ (or $L=S-1$), while the electron left in the final 
state has ${\cal L}=0$ (or $L=S$).
For bound states coupled to the surrounding 2DEG (or to any QW 
imperfections that break the translational symmetry) these selection 
rules are only approximate, and the strength of the optical 
transitions from otherwise non-radiative states is a measure of 
the distortion of these states due to their coupling to the 2DEG.
We have shown\cite{x-pl} that the $e$--$X^-$ Laughlin correlations 
limit high energy $e$--$X^-$ collisions in dilute ($\nu\le{1\over3}$) 
systems, and thus that the approximate selection rules remain valid 
for the $X^-$ states formed in the 2DEG.

Yet another set of selection rules are associated with the electron 
and hole spin degrees of freedom.
If the absence of the mixing of valence subbands, the total electron
and heavy hole spins, $J$ and $J_h$, and projections, $J_z$ and 
$J_{zh}$, are all conserved by $H$.
The recombination events must obey $\Delta J_z=\mp{1\over2}$ and 
$\Delta J_{zh}=\pm{3\over2}$, and the two types of transitions with 
$\Delta(J_z+J_{zh})=\pm1$ correspond to two different polarizations 
of emitted light.
In the presence of the valence subband mixing, the spin of the hole 
is coupled to the hole orbital angular momentum and, through the 
Coulomb interaction, to the orbital angular momentum of the electron.
We assume here that the subband mixing can be neglected and that all 
electron and hole spins are polarized by a large Zeeman energy, so that 
the spin selection rules are always obeyed.

\section{Photoluminescence of Neutral and Charged Excitons
         at Small Layer Separation}
\label{secV}
%%%%%%%%%%%%%%%%%%%%%%%%%%%%%%%%%%%%%%%%%%%%%%%%%%%%%%%%%%%%%%%%%%%%%%
%%%%%%%%%%%%%%%%%%%%%%%%%%%%%%%%%%%%%%%%%%%%%%%%%%%%%%%%%%%%%%%%%%%%%%
\subsection{Laughlin Correlated $e$--$X^-$ Liquid}
\label{secV_1}
%%%%%%%%%%%%%%%%%%%%%%%%%%%%%%%%%%%%%%%%%%%%%%%%%%%%%%%%%%%%%%%%%%%%%%
In narrow QW's ($w\le20$~nm), the $X$'s decouple and the $X^-$ with 
the remaining electrons form a two-component incompressible fluid with 
Laughlin $e$--$X^-$ correlations.\cite{x-fqhe,x-cf} 
Laughlin correlations mean that a number of $e$--$X^-$ pair eigenstates 
that correspond to the smallest average $e$--$X^-$ separation (on a 
sphere, these are the states with maximum $L$; on a plane, these are 
the ones with minimum relative angular momentum) are completely avoided.
\cite{parentage}
The avoiding of a number ($m_{eX^-}$) of highly repulsive $e$--$X^-$ 
pair states is described by a Jastrow prefactor $\prod_{ij}(z_e^{(i)}
-z_{X^-}^{(j)})^{m_{eX^-}}$ in the wavefunction (which leads to a 
generalized, two-component composite fermion model\cite{x-cf}).
This is equivalent to saying that the high energy $e$--$X^-$ 
collisions do not occur, and that the $X^-$'s are effectively 
isolated from the 2DEG.
The isolation of the $X^-$ states is even enhanced at $d>0$ where 
the perpendicular dipole moment of bound $e$--$h$ states increases
their repulsion from one another and from electrons.

Because of the isolation of the $X^-$ states, their binding energies 
and oscillator strengths remain almost unaffected by the presence of 
the surrounding 2DEG.
This is a somewhat surprising result, and one might rather expect 
that the interaction of an $X^-$ with Laughlin quasiparticles could 
affect its recombination.
For example, since the $X^-$--QE or $X^-$--QH scattering breaks the
$\Delta{\cal M}=\Delta{\cal K}=0$ selection rule of an isolated $X^-$,
one might expect the ($\nu$-dependent) recombination of the $X^-_{td}$ 
state embedded in a 2DEG.
The exact numerical calculations for finite $Ne$--$h$ systems with 
$N\le9$ show\cite{bilayer-e} that the $X^-_{td}$ repels QE's and 
attracts QH's.
Although this might suggest discontinuous behavior of PL at Laughlin 
fillings, we find that the oscillator strength of the $X^-_{td}$ remains 
negligible compared to the excitonic emission at any $\nu$ in the whole 
range of $d$ in which it is bound.

\begin{figure}[t]
\epsfxsize=3.40in
\epsffile{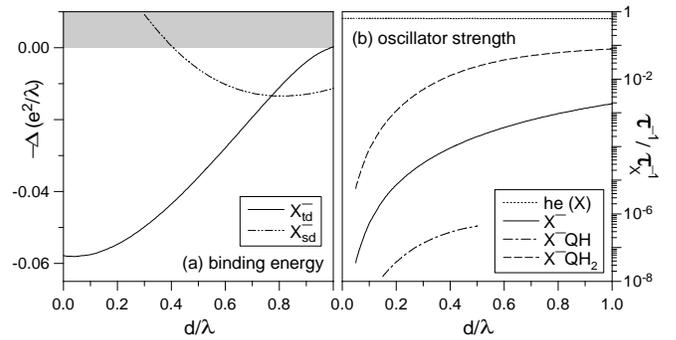}
\caption{
   (a) The binding energy $\Delta$ of the isolated dark triplet 
   ($X^-_{td}$) and dark singlet ($X^-_{sd}$) charged excitons as 
   a function of layer separation $d$ and calculated on a Haldane 
   sphere at $2S=60$. 
   (b) The PL oscillator strength $\tau^{-1}$ of a charged exciton 
   state $X^-_{td}$ binding up to two Laughlin quasiholes QH of 
   the $6e$--$X^-$ incompressible fluid, as a function of $d$ and 
   calculated for an $8e$--$h$ system on a Haldane sphere.
   The $he$ state contains an exciton and originates from the 
   multiplicative state at $d=0$.}
\label{fig2}
\end{figure}

For example, at $\nu>{1\over3}$ all bound $X^-$ states keep far away 
from QE's, and the correlations in the vicinity of each $X^-$ are given 
precisely by the two-component Laughlin wavefunction\cite{halperin} 
$[m_{ee}m_{X^-X^-}m_{eX^-}]$ with Jastrow exponents $m_{ee}=3$ and 
$m_{eX^-}=2$ (the value of $m_{X^-X^-}$ is irrelevant at small $X^-$ 
density).\cite{x-cf,x-pl,bilayer-e}
The oscillator strength of an $X^-$ ``locked'' in such (locally) 
incompressible state [see the $L=2$ ground state in the $9e$--$h$ 
spectrum in Fig.~\ref{fig1}(a)] increases very slowly as a function 
of $d$ and remains negligible compared to the excitonic emission until 
$X^-$ unbinds at $d\approx\lambda$.
At $\nu<{1\over3}$, the $X^-$ can bind one or two QH's to form a new
bound complex $X^-$QH or $X^-$QH$_2$ [see e.g. the $X^-$QH state at 
$L=1$ in the $9e$--$h$ spectrum in Fig.~\ref{fig1}(b)].
Both these complexes have negligible oscillator strength compared to
an exciton.
The reason why binding of one or two QH's to an $X^-$ state does not 
strongly affect its recombination appears to be that the binding of 
QH's means separation of the $X^-$ from neighboring electrons by an 
additional (compared to that of a Laughlin state) charge depletion 
region, without disturbing the $X^-$ state itself.
This only weakly modifies the electron wavefunction in the vicinity 
of the hole which is probed by PL (PL can be regarded as a one electron 
Green function describing the removal of an electron from a state 
initially occupied by a valence hole).
The dependence of the PL oscillator strength of the $X^-$, $X^-$QH, 
and $X^-$QH$_2$ states on $d$ (calculated for the $8e$--$h$ system) 
has been compared to the excitonic emission in Fig.~\ref{fig2}(b).
In Fig.~\ref{fig2}(a) we have also plotted the binding energy of 
an isolated $X^-_{td}$ and compared it to that of a dark singlet 
$X^-_{sd}$.

Summarizing, the PL of a 2DEG in the ``strong coupling'' regime 
(small $d$) occurs from a number of competing radiative bound states: 
$X$, $X^-_s$, and $X^-_{tb}$, whose optical properties are rather 
insensitive to the presence (or density) of the surrounding 2DEG.
Which of these bound states occur in the 2DEG (and their relative 
numbers) depends on their binding energies (which in turn depend 
on $B$, $w$, $w_h$, and $d$) and on their characteristic formation 
($e+X\leftrightarrow X^-$) and recombination times.
The $X^-_{td}$ state remains dark, and no other bound states than 
bright $X$, $X^-_s$, and $X^-_{tb}$, and dark $X^-_{td}$ occur at 
any $\nu$ or $d$.
It is noteworthy that the PL spectrum at small $d$ does not probe 
the interaction of $X$ or $X^-$ states with the 2DEG (at least at 
filling factors up to $\nu\sim{1\over3}$).
As a result, no information about the original correlations of the 
2DEG (before it is perturbed by optically injected valence holes)
can be obtained in a PL experiment in the strong coupling regime.
Indeed, the experimental spectra of symmetrically doped QW's are 
rather insensitive to $\nu$ and show no features at the filling 
factors such as $\nu={1\over3}$ or ${2\over3}$, at which the 
Laughlin--Jain incompressible fluid states of the 2DEG occur 
and the FQH effect is observed in transport experiments.

\subsection{Uncorrelated $e$--$X^-$ System}
\label{secV_2}
%%%%%%%%%%%%%%%%%%%%%%%%%%%%%%%%%%%%%%%%%%%%%%%%%%%%%%%%%%%%%%%%%%%%%%
The $d$-dependence of the energy spectrum of a $3e$--$h$ system (the 
simplest system in which to study interaction of $X^-$ states with 
electrons) shows another interesting feature that might have 
consequence on PL.
At $d=0$, the lowest $3e$--$h$ states describe $e$--$X^-$ pairs (where 
$X^-$ is any of the $X^-_s$, $X^-_{td}$, or $X^-_{tb}$ bound states),
\cite{x-cf,x-pl} and the dependence of energy $E$ on angular momentum 
$L$ for these states is (up to the appropriate $X^-$ binding energy) 
equal to the $e$--$X^-$ interaction pseudopotential $V_{eX^-}(L)$, 
defined\cite{parentage,haldane1} as the dependence of the pair the 
interaction energy on the pair angular momentum.
Due to the dipole--dipole $e$--$X$ repulsion within an $X^-$ at $d>0$, 
the $e$--$X^-$ energies anti-cross the energies of the $2e$--$X$ states
(at the same $L$), in which a $k=0$ exciton is almost decoupled from
two interacting electrons (the states that evolve from the MP states 
at $d=0$).
Because the crossings at larger $L$ (i.e., larger pair energy and 
smaller average $e$--$X^-$ or $e$--$e$ separation) occur at smaller 
$d$, the stability of the $X^-$ in a $e$--$X^-$ collision depends 
critically on both $d$ and $L$.
As we argued in the preceeding section, high energy (i.e., high $L$) 
$e$--$X^-$ collisions do not occur in a Laughlin correlated system.
However, if Laughlin correlations were weakened or destroyed by finite 
QW width ($w>20$~nm), large electron density ($\nu>{1\over3}$), or 
temperature, such collisions could for example result in the break-up 
of otherwise long-lived $X^-$ states ($e+X^-\rightarrow2e+X$) and/or 
their collision-assisted PL from metastable $e$--$X^-_{td}$ pair states 
(which would then occur at a higher energy than the excitonic 
recombination).

\section{Photoluminescence of Fractionally Charged Excitons
         at Large Layer Separation}
\label{secVI}
%%%%%%%%%%%%%%%%%%%%%%%%%%%%%%%%%%%%%%%%%%%%%%%%%%%%%%%%%%%%%%%%%%%%%%
%%%%%%%%%%%%%%%%%%%%%%%%%%%%%%%%%%%%%%%%%%%%%%%%%%%%%%%%%%%%%%%%%%%%%%
It was first realized by Chen and Quinn\cite{chen} that at a large 
layer separation $d$, the PL spectrum of the 2DEG near the Laughlin 
filling factor $\nu\approx(2p+1)^{-1}$ (i.e., at low density of 
Laughlin quasiparticles) can be understood in terms of annihilation 
of a well defined number $n$ of QE's ($0\le n\le2p+1$) and/or creation 
of an appropriate number ($2p+1-n$) of QH's.
Independently of the actual average value of $\nu$ (average over 
the entire 2DEG), the recombination probes a finite area of the 2DEG 
(in the vicinity of the annihilated hole) which has the local filling 
factor of $\nu=(2p+1)^{-1}$ plus a specific number $n$ of QE's bound 
to the hole to form a well defined FCX eigenstate $h$QE$_n$.
For the $\nu={1\over3}$ state, four possible recombination events 
involving QE's and QH's are
\begin{equation}
\label{eq4}
   h+n{\rm QE}\rightarrow (3-n){\rm QH}+\gamma, 
\end{equation}
where $n=0$, 1, 2, or 3, and $\gamma$ denotes the emitted photon.
We have verified this conjecture numerically for the $\nu={1\over3}$ 
state of up to nine electrons.
Indeed, if only the ``first-order'' process (\ref{eq4}) is allowed, 
it describes almost all of the total PL oscillator strength of an 
initial state $h$QE$_n$.
However, we find that this process is allowed only for some of the
$h$QE$_n$ complexes because of the translational symmetry of the 
2DEG (in the vicinity of the position of the recombination event).
As a result of this symmetry, two angular momentum quantum numbers, 
${\cal M}$ and ${\cal K}$, must be simultaneously conserved in PL.
\cite{avron,dzyubenko2}
To study the selection rules following from the (local) 2D translational 
invariance, it is more convenient to use spherical geometry, in which 
they take a simpler form of the conservation of $L$ and $L_z$.\cite{x-pl}

Let us analyze the four processes (\ref{eq4}) in detail.
The emission energy $\omega$ (we set $\hbar=1$) is measured from the 
exciton energy $E_X$ (recombination energy of a free $k=0$ exciton 
in the absence of the 2DEG) at the same $d$. 
The PL intensity of the process $i\rightarrow f+\gamma$ is defined as 
\begin{equation}
   \tau^{-1}=|\left<f|{\cal P}|i\right>|^2,
\end{equation}
so that $\tau^{-1}\equiv1$ for the free-exciton recombination.
Because of the boson--fermion mapping,\cite{canright} identical 
selection rules are obtained using either statistics to describe 
Laughlin quasiparticles.
In the fermionic picture,\cite{hierarchy} the angular momenta of 
a QE in the initial $Ne$--$h$ state $i$ and of a QH in the final 
$(N-1)e$ state $f$ (both at the same monopole strength $2S$) are 
equal, $l_{\rm QE}=l_{\rm QH}=S-N+2$ (but a QE in state $f$ has 
different angular momentum of $S-N+3$).
The hole angular momentum in the initial state is $l_h=S$.

\paragraph*{\rm $h\rightarrow3$QH$+\gamma$:}
%......................................................................
An infinite planar system without any QE's in the vicinity of the 
hole is (locally) represented by a finite spherical system at 
$2S=3(N-1)$.
This gives $l_h=S={3\over2}(N-1)$ and $l_{\rm QH}={1\over2}(N+1)$.
The allowed total angular momenta of three QH's in the final state 
are obtained by addition of three angular momenta $l_{\rm QH}$ 
(of three identical fermionic QH's).
The QH$_3$ molecule (most tightly packed three-QH droplet) 
has $l_{{\rm QH}_3}=l_{\rm QH}+(l_{\rm QH}-1)+(l_{\rm QH}-2)
={3\over2}(N-1)$.
Since $l_h=l_{{\rm QH}_3}$, the $h\rightarrow3$QH$+\gamma$ optical 
process is allowed and creates the QH$_3$ molecule.
It is expected to have rather small intensity $\tau^{-1}$, because 
the ``optical hole'' (vacancy) created in the 2DEG by annihilation 
of a valence hole is given by the single-particle wavefunction 
$\left|m\right>$ of characteristic radius $\lambda$ and has small 
overlap with the much larger QH$_3$ molecule.
Also, the emission energy $\omega$ will be low because of the high 
energy of QH--QH repulsion in the final state (QH$_3$ is the eigenstate 
of pair angular momentum with ${\cal R}=1$, i.e. maximum QH--QH 
repulsion\cite{hierarchy}).

\paragraph*{\rm $h+$QE$\,\rightarrow2$QH$+\gamma$:}
%......................................................................
One QE in the initial state occurs at $2S=3(N-1)-1$ which gives 
$l_h={3\over2}N-2$ and $l_{\rm QE}=l_{\rm QH}={1\over2}N$.
The $h$--QE pair states have angular momentum $L_i$ given by $l_h-
l_{\rm QE}\le L_i\le l_h+l_{\rm QE}$.
The state at $l_{h{\rm QE}}=l_h-l_{\rm QE}=N-2$ describes the $h$QE 
complex with the smallest average $h$--QE separation.
The two QH's in the final state can have pair angular momenta of 
$L_f=2l_{\rm QH}-{\cal R}=N-{\cal R}$ where ${\cal R}$ is an odd 
integer, and the QH$_2$ molecule has $l_{{\rm QH}_2}=N-1$.
Clearly, $l_{h{\rm QE}}\ne L_f$ for any final two-QH state so that
the $h$QE$\,\rightarrow2$QH$+\gamma$ optical process is forbidden.
The $h$QE can only recombine through a ``second-order'' process,
$h$QE$\,\rightarrow3$QH$+$QE$+\gamma$, which will have very small 
intensity.
The only state of an $h$--QE pair that has $L_i=L_f$ and thus can 
recombine through a ``first-order'' process (\ref{eq4}) is the one 
with the next larger value of angular momentum, $l_{h{\rm QE}^*}=N-1$.
This state (denoted by $h$QE*) is\cite{bilayer-e} the first excited 
$h$--QE pair state at $d$ larger than about $\lambda$.
The $h$QE* state may occur at a finite temperature as a result of 
excitation of the long-lived $h$QE complex.
Because QH$_2$ is smaller and has (three times) smaller QH--QH 
repulsion energy than QH$_3$, the $h$QE* is expected to recombine 
with higher intensity and at higher energy than an uncorrelated hole.

\paragraph*{\rm $h+2$QE$\,\rightarrow$QH$+\gamma$:}
%......................................................................
Two QE's in the initial state occur at $2S=3(N-1)-2$ which gives
$l_h={1\over2}(3N-5)$ and $l_{\rm QE}=l_{\rm QH}={1\over2}(N-1)$.
The QE$_2$ molecule has $l_{{\rm QE}_2}=2l_{\rm QE}-1=N-2$.
The $h$--QE$_2$ pair states have $L_i$ given by $l_h-l_{{\rm QE}_2}
\le L_i\le l_h+l_{{\rm QE}_2}$, and the $h$QE$_2$ ground state has 
$l_{h{\rm QE}_2}=l_h-l_{{\rm QE}_2}={1\over2}(N-1)$.
Since $l_{h{\rm QE}_2}=l_{\rm QH}$, the $h$QE$_2$ state is optically
active.
Because of the small size and energy of a single QH, the $h$QE$_2$
will recombine at even higher intensity and higher energy than $h$QE*.

\paragraph*{\rm $h+3$QE$\,\rightarrow\gamma$:}
%......................................................................
Three QE's in the initial state occur at $2S=3(N-1)-3$ which gives
$l_h={3\over2}N-3$ and $l_{\rm QE}={1\over2}N-1$.
The QE$_3$ molecule has $l_{{\rm QE}_3}=l_{\rm QE}+(l_{\rm QE}-1)+
(l_{\rm QE}-2)={3\over2}N-6$.
The $h$--QE$_3$ pair states have $L_i$ given by $l_h-l_{{\rm QE}_3}
\le L_i\le l_h+l_{{\rm QE}_3}$, i.e. $L_i\ge3$.
The smallest value, $l_{h{\rm QE}_3}=3$, describes the $h$QE$_3$ 
molecule, and all other $h+3$QE states (not only the $h$--QE$_3$ 
pair states) have $L_i>3$.
Since $L_f=0$ and $L_i\ge 3$, neither the $h$QE$_3$ state nor its
excitations can recombine through a ``first-order'' process (\ref{eq4}).
Instead, the $h$QE$_3$ recombination must occur through a ``second-order'' 
process, $h+3$QE$_3\rightarrow$QE$+$QH$+\gamma$, which corresponds to 
recombination of an optically active $h$QE$_2$ in the presence of the 
nearby third QE.
This turns out to be allowed only for $L_i>3$, and hence $h$QE$_3$ 
is not only unstable\cite{bilayer-e} but non-radiative as well.
%......................................................................

\subsection{Binding Energy and Optical Properties 
            of $h$QE$_n$ Complexes
            Uncoupled From Charge Excitations of 2DEG}
\label{secVI_1}
%%%%%%%%%%%%%%%%%%%%%%%%%%%%%%%%%%%%%%%%%%%%%%%%%%%%%%%%%%%%%%%%%%%%%%
In order to calculate the binding energies $\Delta$, PL energies 
$\omega$ and oscillator strengths $\tau^{-1}$ of different $h$QE$_n$ 
complexes, a finite $Ne$--$h$ system is diagonalized at the monopole 
strength $2S=3(N-1)-n$, at which $n$ QE's occur in the $\nu={1\over3}$ 
state of $N$ electrons.
In this section, the properties of ``isolated'' $h$QE$_n$ complexes
are studied.
By an isolated $h$QE$_n$ complex we mean one that is uncoupled from 
additional (other then $n$ QE's) charge excitations of the 2DEG, 
that is whose wavefunction involves only the positions of the hole
and of $n$ bound QE's.
The coupling of the $h$QE$_n$ particles to the underlying 2DEG, 
as well as its effect on their binding energy and optical properties,
will be discussed in Sec.~\ref{secVI_2}.

\begin{figure}[t]
\epsfxsize=3.40in
\epsffile{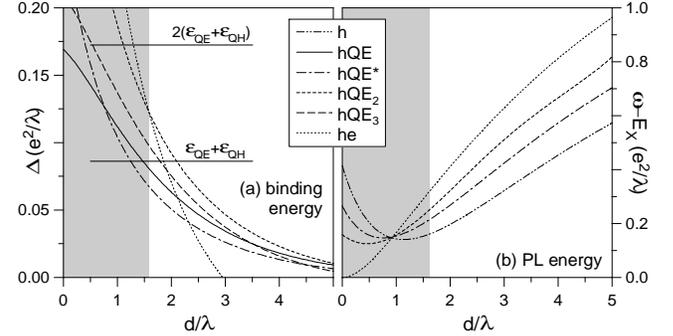}
\caption{
   The binding energy $\Delta$ (a) and recombination energy $\omega$ 
   (b) of fractionally charged excitons $h$QE$_n$ as a function of 
   layer separation $d$, calculated for the $8e$--$h$ system with
   a fixed number of Laughlin quasiparticles in the $8e$ electron
   system ($\epsilon\gg1$; see text).
   $E_X$ is the exciton energy and $\lambda$ is the magnetic length. 
   The $he$ state contains an exciton and originates from the 
   multiplicative states at $d=0$.
   In the shaded parts of both graphs, the $he$ has the largest 
   binding energy and the $h$QE$_n$ complexes do not form.}
\label{fig3}
\end{figure}

To assure that the interaction between the hole and the 2DEG is weak 
compared to the energy $\varepsilon_{\rm QE}+\varepsilon_{\rm QH}$ 
($\approx0.1\,e^2/\lambda$ for an infinite system) needed to create 
additional QE--QH pairs in the 2DEG, the charge of the hole is set 
to $e/\epsilon$ where $\epsilon\gg1$.
This guarantees that the lowest $Ne$--$h$ states contain exactly 
$n$ QE's interacting with the hole (even if the large $h$--QE
attraction at a finite $d$ and $\epsilon=1$ induced additional QE--QH 
pair excitations to screen the hole with additional QE's) and thus 
that the ground state is the $h$QE$_n$ bound state.
If $\epsilon$ is sufficiently large, $\Delta$, $\omega$, and 
$\tau^{-1}$ calculated in this way are independent of $\epsilon$ 
and describe the ``ideal'' $h$QE$_n$ wavefunctions, in which a hole 
is bound to a QE$_n$ molecule [the $n$QE state with the maximum 
angular momentum $l_{{\rm QE}_n}=nl_{\rm QE}-{1\over2}n(n-1)$].
The PL intensities $\tau^{-1}$ are also independent of $d$, and the 
values calculated for the $Ne$--$h$ systems with $N\le9$ are listed 
in Tab.~\ref{tab1}.
\begin{table}
\caption{
   The PL oscillator strength $\tau^{-1}_N$ (in the units of the 
   oscillator strength of a free $k=0$ exciton) of fractionally 
   charged excitons $h$QE$_n$ calculated in the $Ne$--$h$ systems 
   ($6\le N\le9$) on a Haldane sphere.}
\begin{tabular}{rcccccc}
   &MP&$h$&$h$QE&$h$QE*&$h$QE$_2$&$h$QE$_3$\\
\tableline
   $\tau^{-1}_6$&0.6154&0.0231&---&0.0968&0.1144&---\\
   $\tau^{-1}_7$&0.6250&0.0187&---&0.0767&0.0938&---\\
   $\tau^{-1}_8$&0.6316&0.0160&---&0.0649&0.0791&---\\
   $\tau^{-1}_9$&0.6364&0.0138&---&0.0556&0.0680&--- 
\end{tabular}
\label{tab1}
\end{table}
To obtain the dependence of binding energies $\Delta$ and PL energies 
$\omega$ on $d$, the $h$--QE attraction is multiplied by $\epsilon$.
The data obtained for $N=8$ are plotted in Fig.~\ref{fig3}.

The MP state in Tab.~\ref{tab1} is the lowest energy $L=0$ state at 
$d=0$ and $2S=3(N-2)$, in which the $k=0$ exciton is decoupled from
the $L=0$ Laughlin state of $N-1$ electrons.
Its PL oscillator strength equals 
\begin{equation}
\label{eq5}
   \tau_{\rm MP}^{-1}=1-{N-1\over2S+1}\rightarrow1-\nu 
\end{equation}
for $N\rightarrow\infty$.
The $he$ state in Fig.~\ref{fig3}(b) is the state that evolves from 
this MP state when $d$ is increased (it is calculated with full hole 
charge, $\epsilon=1$), and it has been identified in the $9e$--$h$ 
spectrum at $d=0.25\lambda$ in Fig.~\ref{fig1}(b).
Its PL intensity is almost constant at small $d$ (when $d$ increases 
from 0 to 1, 1.5, and $2\lambda$, then $\tau_{he}^{-1}$ decreases by 
1\%, 6\%, and 14\%, respectively).
Constant $\tau_{he}^{-1}$ means almost unchanged wavefunction, and 
thus the $he$ state contains a $k=0$ exciton that is only weakly 
distorted due to interaction with the 2DEG.
At $d>2\lambda$ the $e$--$e$ correlations become dominant and the $he$ 
state undergoes complete reconstruction ($\tau_{he}^{-1}$ drops quickly 
and $\Delta_{he}$ becomes negative).
No excitonic recombination is expected in PL spectra at $d$ much larger 
than $2\lambda$.
At $d=0$, the PL energy $\omega_{he}$ of the $he$ state equals the 
energy $E_X$ of a single exciton (because of the hidden symmetry).
At $d>0$, $\omega_{he}>E_X$ because the dipole moment of the $k=0$ 
exciton (perpendicular to the layers) causes its repulsion by the 
surrounding electrons.

For FCX's, both intensity and energy behave as predicted in proceeding 
paragraphs.
The occurrence of four possible PL peaks (although not all of them will 
occur at the same $d$ due to different ranges of stability of different
complexes; see Sec.~\ref{secVII}) reflects quantization of the total 
charge $-q$ that can be bound to a hole in the units of the charge of 
Laughlin quasiparticles: $q/e=1$, ${2\over3}$, ${1\over3}$, and 0 for 
$he$, $h$QE$_2$, $h$QE*, and $h$ states, respectively ($q$ does not 
include the uniform charge density of the underlying Laughlin state).

At $d>\lambda$ all radiative FCX's emit at the energy below 
$\omega_{he}$.
The ordering, $\omega_h<\omega_{h{\rm QE}^*}<\omega_{h{\rm QE}_2}<
\omega_{he}$, and almost equal spacing between the PL energies at 
$d>\lambda$ results from the comparison of the initial and final 
state energies, 
\begin{eqnarray}
   E_i&=&
   N\varepsilon_0+n\varepsilon_{\rm QE}+{n(n-1)\over2}V_{\rm QE}
   -\Delta_{h{\rm QE}_n},
\nonumber\\
   E_f&=&
   (N-1)\varepsilon_0+(3-n)\varepsilon_{\rm QH}+{(3-n)(2-n)\over2}
   V_{\rm QH},
\end{eqnarray}
where $\varepsilon_0$ is the Laughlin ground state energy per electron, 
and $V_{\rm QE}=V_{\rm QE-QE}(1)$ and $V_{\rm QH}=V_{\rm QH-QH}(1)$ 
are the energies of QE--QE and QH--QH interactions per pair.
At $d\gg\lambda$, when the $h$QE$_n$ binding energy can be neglected, 
for the separations between the three FCX peaks we obtain
\begin{eqnarray}
   \omega_{h{\rm QE}^*}-\omega_h&=&
   \varepsilon_{\rm QE}+\varepsilon_{\rm QH}+2V_{\rm QH},
\nonumber\\
   \omega_{h{\rm QE}_2}-\omega_{h{\rm QE}^*}&=&
   \varepsilon_{\rm QE}+\varepsilon_{\rm QH}+V_{\rm QH}+V_{\rm QE}.
\end{eqnarray}
At smaller $d$, the separation between peaks decreases because
$\Delta_{h{\rm QE}_2}>\Delta_{h{\rm QE}*}>\Delta_h=0$.
The crossing occurs at $d\approx\lambda$.
At $d<\lambda$ the ordering of the PL energies in Fig.~\ref{fig3}(b)
is reversed, but this (shaded) part of the graph has no physical 
significance (FCX's do not occur).
Two other points could be important:

Firstly, the PL oscillator strengths of all radiative FCX's in 
Tab.~\ref{tab1} ($h$, $h$QE*, and $h$QE$_2$) decrease as a function 
of $N$.
Hence, the results of our finite size calculations alone are not 
conclusive as to whether the recombination of these complexes 
contributes to the PL spectra of infinite systems.
However, the vanishing of $\tau_h^{-1}$, $\tau_{h{\rm QE}^*}^{-1}$, 
and $\tau_{h{\rm QE}_2}^{-1}$ for $N\rightarrow\infty$ would have to 
result from an additional, unexpected symmetry recovered in this limit 
(in analogy to the 2D translational/rotational symmetry which resulted 
in vanishing of $\tau_{h{\rm QE}}^{-1}$ and $\tau_{h{\rm QE}_3}^{-1}$).
Therefore, it is most likely that $h$, $h$QE*, and $h$QE$_2$ remain
(weakly) optically active in an infinite system, and our data suggests 
that $\tau_h^{-1}<\tau_{h{\rm QE}^*}^{-1}<\tau_{h{\rm QE}_2}^{-1}$.

And secondly, Eq.~(\ref{eq5}) implies $\tau_{\rm MP}^{-1}\rightarrow0$ 
for $\nu\rightarrow1$, in complete disagreement with experiments which 
show strong excitonic recombination at $\nu=1$ even at the highest 
available magnetic fields.
This means that the description of the experimentally observed excitonic 
recombination in terms of the ``hidden symmetry'' of the lowest LL fails 
completely. 
Since the LL mixing is more important for the excitonic state $he$ than 
for the FCX complexes (due to larger interaction energy compared to the 
cyclotron energy), one can expect enhancement of the $he$ binding at 
finite $B$ compared to the FCX binding energies.
Although this enhancement depends on a particular system ($B$, QW width, 
etc.), we have checked that for parameters of Ref.~\onlinecite{x-pl} 
(symmetric 11.5~nm GaAs QW), inclusion of excited LL's lowers the energy 
of a free exciton by 0.25, 0.12, 0.035, and 0.015~$e^2/\lambda$ at $B=5$, 
10, 30, and 50~T, respectively.
Hence, at high magnetic fields ($B\ge10$~T) it can be assumed that even 
though our $he$ energy obtained in the lowest LL approximation is not 
very accurate, the error of this approximation is smaller than the peak 
splittings in Fig.~\ref{fig3}(b) and the ordering of peaks is predicted 
correctly.
However, at low fields ($B\le5$~T) the excitonic state $he$ will 
probably remain bound up to much larger $d$ than predicted in 
Fig.~\ref{fig3}(a), and its recombination could occur below that 
of FCX complexes.

\subsection{Binding Energy and Optical Properties 
            of $h$QE$_n$ Complexes 
            Coupled To Charge Excitations of 2DEG}
\label{secVI_2}
%%%%%%%%%%%%%%%%%%%%%%%%%%%%%%%%%%%%%%%%%%%%%%%%%%%%%%%%%%%%%%%%%%%%%%
In this section we calculate the optical properties of the $h$QE$_n$ 
complexes coupled to the underlying 2DEG, that is of actual complexes 
that occur in an $e$--$h$ system at a finite $d$.
To do so, the finite $Ne$--$h$ spectra similar to those in 
Fig.~\ref{fig1} are calculated including both $e$--$e$ and $e$--$h$ 
interactions (i.e., with $\epsilon=1$).
The $h$QE$_n$ complexes are identified in these spectra as low energy
states with appropriate angular momentum.
The binding energy $\Delta$, PL recombination energy $\omega$, 
and PL oscillator strength $\tau^{-1}$ are calculated for these 
states and compared with the values obtained for $\epsilon\gg1$ 
in Sec.~\ref{secVI_1}. 
Small difference between the values obtained for $\epsilon\gg1$ and 
$\epsilon=1$, as well as the convergence of the two in the 
$d\rightarrow\infty$ limit, confirms the identification of $h$QE$_n$ 
states in the $Ne$--$h$ spectra.

\begin{figure}[t]
\epsfxsize=3.40in
\epsffile{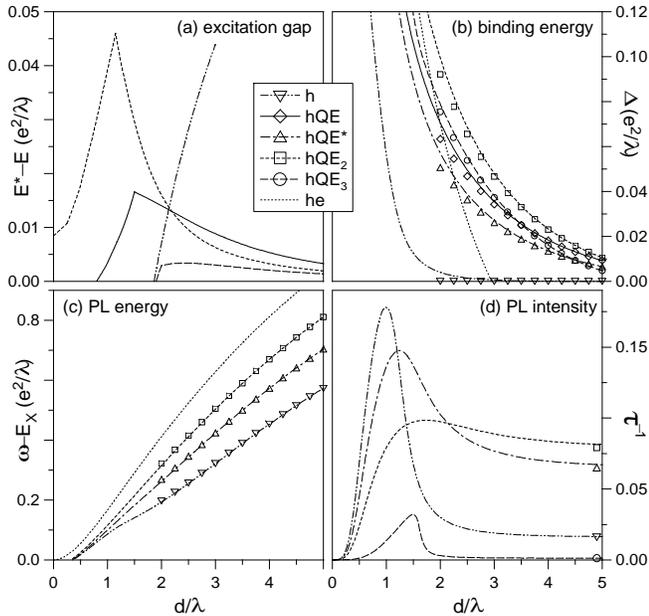}
\caption{
   The excitation gap $E^*-E$ (a), binding energy $\Delta$ (b), 
   recombination energy $\omega$ (c), and oscillator strength 
   $\tau^{-1}$ (d) of fractionally charged excitons $h$QE$_n$ 
   as a function of layer separation $d$, calculated for the 
   $8e$--$h$ system.
   $E_X$ is the exciton energy and $\lambda$ is the magnetic length. 
   The $he$ state contains an exciton and originates from the 
   multiplicative states at $d=0$.}
\label{fig4}
\end{figure}

Fig.~\ref{fig4} shows the data calculated for an $8e$--$h$ system.
We have checked that the curves plotted here for $N=8$ are very close 
to those obtained for $N=7$ or 9, so that all important properties of 
an extended system can be understood from a rather simple $8e$--$h$
computation.
In four frames, for each $h$QE$_n$ we plot: 
(a) the excitation gap $E^*-E$ above the $h$QE$_n$ ground state;
(b) the binding energy $\Delta$;
(c) the recombination energy $\omega$; and
(d) the recombination intensity (PL oscillator strength) $\tau^{-1}$.
The excitation gaps and the recombination energies and intensities 
are obtained from the spectra at $2S=3(N-1)-n$ in which the $h$QE$_n$ 
complexes occur.
The binding energy $\Delta$ is defined in such way that $E_{h{\rm QE}_n}
=E_{{\rm QE}_n}+V_{h-{\rm LS}}-\Delta$, where $E_{h{\rm QE}_n}$ is the 
energy of the $Ne$--$h$ system in state $h$QE$_n$ calculated at 
$2S=3(N-1)-n$, $E_{{\rm QE}_n}$ is the energy of the $Ne$ system in 
state QE$_n$ calculated at the same $2S=3(N-1)-n$, and $V_{h-{\rm LS}}$ 
is the self-energy of the hole in Laughlin $\nu={1\over3}$ ground state 
at $2S=3(N-1)$.
As described in Sec.~\ref{secVI_1}, $V_{h-{\rm LS}}$ is calculated by 
setting the hole charge to a very small fraction of $+e$ so that it 
does not perturb the Laughlin ground state.
The $he$ curves in Fig.~\ref{fig4} are identical to those in 
Fig.~\ref{fig3}.
The PL intensity of the $he$ state (which is the MP state at $d=0$)
is too large (see Tab.~\ref{tab1}) to fit in Fig.~\ref{fig4}(d).

The lines in Fig.~\ref{fig4} show data obtained from the spectra 
similar to those in Fig.~\ref{fig1}, that is including all effects 
of $e$--$h$ interactions.
For comparison, with symbols we have replotted the data from 
Fig.~\ref{fig3} obtained for $\epsilon\gg1$ to assure that, at any 
$d$, the obtained low energy eigenstates are given exactly by the 
$h$QE$_n$ wavefunctions.
At $d>\lambda$, very good agreement between binding energies and PL 
energies calculated for $\epsilon=1$ (lines) and $\epsilon\gg1$ 
(symbols) confirms our identification of $h$QE$_n$ states in low 
energy $Ne$--$h$ spectra.
The PL intensities $\tau^{-1}$ calculated for $\epsilon=1$ (lines) 
converge to those obtained for $\epsilon\gg1$ and listed in 
Tab.~\ref{tab1} (symbols).
The good agreement between the lines and symbols at $d>2\lambda$ shows 
that the $h$QE$_n$ states identified in that $Ne$--$h$ spectra are indeed 
described by exact $h$QE$_n$ wavefunctions.
At $d<\lambda$ the two calculations give quite different results 
which confirms that the description of actual $Ne$--$h$ eigenstates
in terms of the hole interacting with Laughlin quasiparticles of the
2DEG is inappropriate (the correct picture is that of a two-component 
$e$--$X^-$ fluid).
The formation of $h$QE$_n$ complexes at $d$ larger than about 
$1.5\lambda$ can be seen most clearly in the $\tau^{-1}(d)$ curves.
For example, while it is impossible to detect the point of transition 
between the $X^-$QH$_2$ and $h$QE$_2$ complexes in the dependence of 
energy spectrum in Fig.~\ref{fig1} on $d$ (because $l_{X^-{\rm QH}_2}=
l_{h{\rm QE}_2}$), it is clearly visible at $d\approx1.5\lambda$
in Fig.~\ref{fig4}(d).

The analysis of the characteristics of $h$QE$_n$ complexes plotted 
in Figs.~\ref{fig3} and \ref{fig4} leads to the conclusion that the 
bound complex most important for understanding PL in the weak coupling 
regime ($d>1.5\lambda$) is $h$QE$_2$, which has the largest binding 
energy $\Delta$, and significant excitation energy $E^*-E$ and PL 
oscillator strength $\tau^{-1}$.
The $h$QE is also a strongly bound complex with large excitation 
energy, but it is non-radiative (at least, in the absence of scattering 
or disorder).
Although $h$QE is dark, its first excited state, $h$QE*, is both bound 
and radiative and can contribute to the PL spectrum.
The charge neutral ``anyon exciton'' $h$QE$_3$ suggested by Rashba et al.
\cite{rashba} is neither bound nor radiative.
Finally, the radiative excitonic state ($k=0$ charge neutral $e$--$h$ 
pair weakly coupled to the 2DEG) breaks apart at $d>2\lambda$.

%%%%%%%%%%%%%%%%%%%%%%%%%%%%%%%%%%%%%%%%%%%%%%%%%%%%%%%%%%%%%%%%%%%%%%
\section{Stability and Emission of Different Bound States:
         PL Spectra at Different Layer Separations}
\label{secVII}
\begin{figure}[t]
\epsfxsize=3.40in
\epsffile{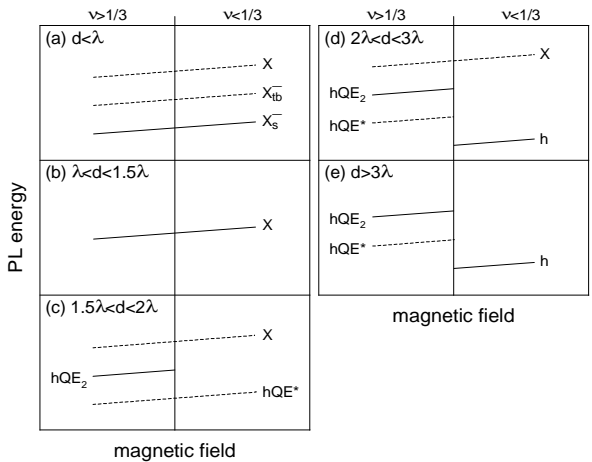}
\caption{
   The schematic PL spectra (PL energy vs.\ magnetic field) 
   near the filling factor $\nu={1\over3}$ at different layer 
   separations $d$.
   Solid and dashed lines mark recombination from ground and 
   excited states, respectively.
   $\lambda$ is the magnetic length. 
}
\label{fig5}
\end{figure}

%%%%%%%%%%%%%%%%%%%%%%%%%%%%%%%%%%%%%%%%%%%%%%%%%%%%%%%%%%%%%%%%%%%%%%
%%%%%%%%%%%%%%%%%%%%%%%%%%%%%%%%%%%%%%%%%%%%%%%%%%%%%%%%%%%%%%%%%%%%%%
The information presented in Figs.~\ref{fig3} and \ref{fig4} and in 
Tab.~\ref{tab1} allows understanding of anomalies observed in the PL 
spectra of the 2DEG near $\nu={1\over3}$.
The crucial observations are:
(i) the most strongly bound complexes at small layer separation $d$ 
are the $k=0$ state of a charge-neutral exciton $X$ and different 
states of charged excitons $X^-$;
(ii) at larger $d$, the most stable complexes are the bright $h$QE$_2$ 
and dark $h$QE (whose weakly excited state $h$QE* is bright);
(iii) no charge-neutral ``dressed exciton'' states at $k\ne0$ occur; 
and
(iv) the charge-neutral ``anyon exciton'' $h$QE$_3$ is neither stable 
nor radiative.
Depending on the layer separation $d$ and on whether $\nu$ is larger 
or smaller than ${1\over3}$, the following behavior is expected 
(see the schematic PL spectra in Fig.~\ref{fig5}; it should be 
understood that the PL spectrum changes continuously as a function 
of $d$ but discontinuously as a function of $\nu$).

\paragraph*{\rm $d<\lambda$:}
%......................................................................
The holes bind one or two ``whole'' electrons to form $k=0$ neutral 
excitons or various charged exciton states (the relative numbers of 
$X^-_s$, $X^-_{td}$, and $X^-_{tb}$ depend on $B$, temperature, etc.).
No ``dressed exciton'' states with $k\ne0$ (in-plane dipole moment) 
occur.
The $k=0$ excitons weakly couple to the 2DEG, and the $X^-$'s are 
effectively isolated from the 2DEG because of Laughlin $e$--$X^-$ 
correlations.
As a result, neither the recombination of a $k=0$ excitons and {\rm 
radiative} $X^-$ states ($X^-_s$ and $X^-_{tb}$), nor the lack of
recombination of the dark $X^-_{td}$ state are significantly affected 
by the 2DEG.
Only the $X^-$'s will occur in the absolute ground state of the system.
However, because the exciton has shorter optical lifetime than all the
$X^-$ states, the PL spectrum at finite temperatures contains peaks 
corresponding to both exciton (in our notation: $he$) and $X^-$ 
recombination. 
At $d>0$, the $he$ recombination energy is larger than the bare 
($\nu=0$) exciton energy $E_X$ at the same $d$ due to the $e$--$X$ 
repulsion.

\paragraph*{\rm $\lambda\le d<1.5\lambda$:}
%......................................................................
The $X^-$'s unbind but the neutral excitons still exist.
The FCX complexes ($h$QE and $h$QE$_2$) also occur, as the QE--QH pairs 
are spontaneously created in the 2DEG to screen the charge of each hole.
However, the exciton has both the largest binding energy and the largest
PL oscillator strength, and its recombination dominates the PL spectrum.
A similar electric-field induced ionization of $X^-$'s in a QW has been 
demonstrated at $B=0$ by Shields et al.\cite{shields2}

\paragraph*{\rm $1.5\lambda\le d<2\lambda$:}
%......................................................................
The excitons still exist but they no longer have maximum binding energy.
To screen the charge of each hole, one QE--QH pair is spontaneously 
created in the 2DEG to form the FCX complex $h$QE ($h\rightarrow h$QE
$+$QH).
Since $h$QE$_2$ has larger binding energy than $h$QE, it can also be 
formed in the presence of excess QE's ($h$QE$+$QE$\,\rightarrow h$QE$_2$), 
but it will be destroyed in the presence of excess QH's ($h$QE$_2+$QH$
\rightarrow h$QE).
Therefore, a discontinuity is expected when $\nu$ crosses ${1\over3}$:
At $\nu>{1\over3}$, the dark $h$QE and the bright $h$QE$_2$ co-exist 
and the $h$QE recombination can occur either through binding of the 
second QE to form a bright $h$QE$_2$ (note that $\nu_{\rm QE}=1$ occurs 
at $\nu={2\over5}$ and thus, except for $\nu$ almost equal to ${1\over3}$,
the QE density is larger than the hole density), or through excitation 
to a bright $h$QE*.
At $\nu<{1\over3}$, the $h$QE is the only stable complex and its dominant 
recombination channel is through excitation to the bright $h$QE* state
which emits at similar rate but lower energy than $h$QE$_2$ (by about
$\varepsilon_{\rm QE}+\varepsilon_{\rm QH}$).
The strongly radiative $k=0$ excitons ($he$) are also visible at finite 
temperatures.
Clearly, different temperature dependence of the emission from the 
ground state $h$QE$_2$ and from the excited states $h$QE* and $he$ 
is expected.

\paragraph*{\rm $2\lambda\le d<3\lambda$:}
%......................................................................
The excitons still exist but they have very small binding energy.
No QE--QH pairs are spontaneously created and the holes can only bind 
existing QE's, which leads to discontinuity when $\nu$ crosses 
${1\over3}$:
At $\nu>{1\over3}$, the relative numbers of $h$QE, $h$QE*, and $h$QE$_2$ 
depend on the hole and QE densities and temperature.
However, because the QE density can be assumed larger than the hole 
density and the $h$QE is long-lived, both $h$QE* and $h$QE$_2$ are 
expected to show in the PL spectrum, emitting at energies different 
by about $\varepsilon_{\rm QE}+\varepsilon_{\rm QH}$.
At $\nu<{1\over3}$, there are no QE's to bind, and the holes repel the 
existing QH's.
In the ground state, there is no response of the 2DEG to the hole, 
whose recombination occurs at the local filling factor $\nu={1\over3}$ 
(and probes the spectral function of an electron annihilated in an 
undisturbed Laughlin $\nu={1\over3}$ state).
Although the optical lifetime of an unbound hole is fairly long, 
no bound radiative FCX's are expected at low temperatures since the 
recombination of $h$QE$_2$ or $h$QE* must occur through the formation 
of an unstable $h$QE ($h$QE$+$QH$\,\rightarrow h$) followed by either 
binding of a second QE to form the $h$QE$_2$ or an excitation to form 
the $h$QE*.
Although weakly bound, the neutral exciton ($he$) has much large 
oscillator strength than an uncorrelated hole, and it might also be 
observed in PL at a finite temperature.
The exciton binding strongly depends on the LL mixing, so it is more
likely to exist at lower $B$ (in the lower density samples). 

\paragraph*{\rm $d\ge3\lambda$:}
%......................................................................
No excitons ($he$) occur, and the recombination can only occur from 
the $h$QE$_2$, $h$QE*, or $h$ states, with a discontinuity at 
$\nu={1\over3}$.

\section{Conclusion}
\label{secVIII}
%%%%%%%%%%%%%%%%%%%%%%%%%%%%%%%%%%%%%%%%%%%%%%%%%%%%%%%%%%%%%%%%%%%%%%
%%%%%%%%%%%%%%%%%%%%%%%%%%%%%%%%%%%%%%%%%%%%%%%%%%%%%%%%%%%%%%%%%%%%%%
We have studied PL from a 2DEG in the fractional quantum Hall regime
as a function of the separation $d$ between the electron and valence 
hole layers.
Possible bound states in which the hole can occur have been identified 
and characterized in terms of such single particle quantities as the 
angular momentum, binding energy, recombination energy, and oscillator 
strength.
The strict optical selection rules for these bound states have been 
formulated, following from the (local) 2D translational symmetry of 
each state.
Only some of the bound states turn out radiative, and their relative 
oscillator strengths are predicted from a rather simple analysis.
The discussion is illustrated with the results of exact numerical 
calculations in Haldane's spherical geometry for a hole interacting 
with up to nine electrons at filling factors $\nu\sim{1\over3}$.

Different response of the 2DEG to the optically injected hole in the
strong and weak coupling regime results in a complete reconstruction 
of the PL spectrum at $d$ of the order of the magnetic length $\lambda$.
At $d<\lambda$, the hole binds one or two electrons to form a neutral 
exciton state $X$ or various charged exciton states $X^-$.
The PL spectrum in this regime depends on the lifetimes and binding 
energies of the $X$ and $X^-$ states, rather than on the original 
correlations of the 2DEG.
No anomaly occurs in PL at the Laughlin filling factor $\nu={1\over3}$, 
at which the FQH effect is observed in transport experiments.

At $d$ larger than about $2\lambda$, the Coulomb potential of the 
distant hole becomes too weak and its range becomes too large to 
bind individual electrons and form the $X$ or $X^-$ states.
Instead, the hole interacts with charge excitations of the 2DEG,
namely, repels QH's and attracts QE's of the Laughlin incompressible
$\nu={1\over3}$ fluid.
The resulting states in which the hole can occur are the uncorrelated 
state $h$ (in which the free hole moves in the rigid electron Laughlin 
fluid at a local filling factor $\nu={1\over3}$), and the fractionally 
charged excitons $h$QE and $h$QE$_2$ (in which the hole binds one or 
two QE's).
Different states have very different optical properties (recombination 
lifetimes and energies), and which of them occur depends critically 
on whether QE's are present in the 2DEG.
Therefore, discontinuities occur in the PL spectrum at $\nu={1\over3}$. 

Our results invalidate two suggestive concepts proposed to understand
the numerical $Ne$--$h$ spectra and the observed PL of a 2DEG.
Firstly, the ``dressed exciton'' states\cite{wang,apalkov} with finite 
momentum ($k\ne0$) do not occur in the low energy spectra of $e$--$h$ 
systems at small $d$.
And secondly, the charge neutral ``anyon exciton'' states\cite{rashba} 
are neither stable nor radiative at any value of $d$.

\section*{Acknowledgment}
\label{secIX}
%%%%%%%%%%%%%%%%%%%%%%%%%%%%%%%%%%%%%%%%%%%%%%%%%%%%%%%%%%%%%%%%%%%%%%
%%%%%%%%%%%%%%%%%%%%%%%%%%%%%%%%%%%%%%%%%%%%%%%%%%%%%%%%%%%%%%%%%%%%%%
The authors acknowledge partial support by the Materials Research 
Program of Basic Energy Sciences, US Department of Energy.
AW acknowledges partial support from the Polish State Committee 
for Scientific Research (KBN) grant 2P03B11118.
The authors thank K. S. Yi (Pusan National University, Korea) who
participated in the early stages of this study, and P. Hawrylak 
(National Research Council, Canada) and M. Potemski (High Magnetic
Field Laboratory, Grenoble, France) for helpful discussions.

\end{document}